\documentclass[seceq,supplement]{ptptex}
\usepackage{wrapft}
\usepackage{graphicx}

\pubinfo{Vol.~1XX, No.~X, Xxxx 200X}

\title{
Non-Universal Finite Size Effects \\
with Universal Infinite-Size Free Energy for the $\alpha$-XY model
}

\author{Shin-itiro \textsc{Goto} and Yoshiyuki Y. \textsc{Yamaguchi}}

\inst{Department of Applied Mathematics and Physics, Kyoto University, Kyoto 606-8501, Japan}

\recdate{
Xxxx X, 200X}

\abst{
We study finite size effects
  in a family of systems 
  in which a parameter controls interaction-range.  
  In the long-range regime where the infinite-size free energy 
  is universal,
  we show that the finite size effects are not universal
  but depend on the interaction-range.
  The finite size effects are observed through 
  discrepancies
  between time-averages of macroscopic variables in Hamiltonian dynamics
  and canonical averages of ones
  with infinite degrees of freedom.
  For a high energy  regime, 
  the relation to a pair of the discrepancies
  is theoretically predicted and numerically confirmed.
We also numerically show that
the finite-size effects of macroscopic variables 
in the canonical ensemble  
are close to ones in the dynamical systems. 
}

\begin{document}
\newcommand{\average}[1]{\left\langle #1 \right\rangle}
\newcommand{\non}{\nonumber}
\newcommand{\beq}{\begin{equation}}
\newcommand{\beqa}{\begin{eqnarray}}
\newcommand{\eeq}{\end{equation}}
\newcommand{\eeqa}{\end{eqnarray}}
\newcommand{\lb}{\label}
\newcommand{\fr}[1]{(\ref{#1})}

\maketitle

\section{Introduction}
In many body systems, 
the range of interactions governs physical features.
Long-range interactions induce a phase transition,
and typical ones are the mean-field interactions.
For instance, the Hamiltonian mean-field (HMF) model
which consists of planar rotators
shows a second-order phase transition
between a low energy ordered phase and a high energy disordered phase
\cite{antoni-95}.
On the other hand, short-range interactions 
are hard to occur the phase transition, 
particularly in low-dimensional systems. If the planar rotators are on a 
one-dimensional lattice with nearest-neighbor interactions, 
then no phase transition occurs.
 
To investigate how the range of interactions affects
the existence of a phase transition,
the HMF model is extended to a model having algebraically decaying
interactions, called the $\alpha$-XY model,
described by the following Hamiltonian \cite{anteneodo-98}:
\begin{equation}
  \label{eq:alpha-XY}
  H = \dfrac{1}{2} \sum_{i=1}^{N} p_{i}^{2}
  + \dfrac{1}{2\tilde{N}} \sum_{i=1}^{N} \sum_{j=1}^{N}
  \dfrac{ 1-\cos(q_{i}-q_{j}) }{ r_{ij}^{\alpha} }.
\end{equation}
Each rotator is placed on a site of the 
periodic $d$-dimensional simple lattice,
that is $N=L^{d}$ for the lattice of size $L$,
and $r_{ij}$ is the shortest distance between the sites $i$ and $j$.
The $\alpha$-XY model includes the HMF model as $\alpha=0$,
and the nearest neighbor interaction realizes
in $\alpha\to\infty$. 
The factor $\tilde{N}$ 
is introduced for recovering
the extensivity of the system, 
but the factor does not make the system additive for $0\leq\alpha/d<1$.
In this article, 
$\tilde{N}$ is defined as $\sum_{i=1}^N1/r_{ij}^{\alpha}$
\cite{campa-00}.
In the long-range regime $0\leq\alpha/d<1$,
Campa {\it et al.} show that the $\alpha$-XY model
including the external magnetic field 
essentially has the same canonical partition function 
with the HMF model in $N\to\infty$\cite{campa-00},
and the critical value of the energy for a second-order phase transition
is common as $U=E/N=0.75$, with $E$ being the value of the Hamiltonian.
Recent studies of the $\alpha$-XY model are briefly summarized
in Ref.\citen{giansanti-02}.

In this article, in  the long-range regime,
we investigate how macroscopic variables for 
the $\alpha$-XY system with finite $N$
approaches the thermodynamic limit ($N\to\infty$).
They could go towards canonical averages as $N$ increases
(see Appendix, and Refs.~\citen{latora-99,yamaguchi-03,yamaguchi-04} for the HMF model),
but it is not obvious 
whether the convergence speed depends on $\alpha/d$ or not.
Some non-universalities appear as dynamical aspects in $0\leq\alpha/d<1$ 
\cite{giansanti-02},
and hence we compare canonical averages of the macroscopic variables
in $N\to\infty$ and time-averaged variables with finite $N$.
The time-averaged variables are 
obtained by integrating the canonical equation of motion
yielded from the Hamiltonian (\ref{eq:alpha-XY}).
We set $d=1$ for the convenience of numerical computations,
and show 
that, for the supercritical energy regime,
the time-averages of macroscopic variables
with finite $N$ algebraically approach the canonical averages
taken in $N\to\infty$ as $N$ increases,
and the exponents depend on $\alpha$.

\section{Theoretical Prediction}
As macroscopic variables we observe,
we adopt modulus of the magnetization and the temperature.
The magnetization vector $\vec{M}:=(M_{x},M_{y})$ 
and its modulus $M$ are defined as
$\vec{M}:= \sum_{i=1}^{N} (\cos q_{i}, \sin q_{i} )/N$ and 
$M := \sqrt{M_{x}^{2}+M_{y}^{2}}$,
respectively.
The temperature in the sence of dynamics 
is defined as $T_N^d := \sum_{j=1}^{N} p_{j}^{2}/N$.
Note that we numerically compute averages through Hamiltonian dynamics
instead of canonical ensemble,
then a fixed variable is neither $T$ nor $T_N^d$, 
but the internal energy $U=E/N$.

In this section, we estimate the following two discrepancies
\begin{equation}
  \label{eq:deltas}
  \delta_{M}(N) := \average{M}_{N} - \average{M}_{\infty},
  \quad
  \delta_{T}(N) := T_{N} - T,
\end{equation}
where $\average{X}_N$ and $\average{X}_{\infty}$ denote the canonical 
average of $X$ with finite $N$ degrees of freedom and one in $N\to\infty$ 
respectively. 
$\average{U}_N$ and $\average{M}_N$ are calculated in Ref
\citen{GY-05} with the aid of the saddle-point approximation\cite{campa-00}.
$T_{N}$ is the temperature for the 
system with $N$ degrees of freedom
and is defined as
\begin{equation}
  \label{eq:UandUN}
U = \average{U}_{N}(T_{N}).
\end{equation}
The relation \fr{eq:UandUN} bridges between
Hamiltonian dynamics and canonical statistics.
  We require that the relation \fr{eq:UandUN} holds
  in any degrees of freedom and hence
$
\average{U}_{N}(T_{N}) = \average{U}_{\infty}(T).
$
The average $\average{U}_{\infty}(T)$ is related to 
$T$ and $\average{M}^{2}_{\infty}$ as
\begin{equation}
  \lb{eq:internal-energy}
  2 \average{U}_{\infty}(T) = T + 1 -[\average{M}_{\infty}(T)]^2.
\end{equation}
The relation \fr{eq:internal-energy} is straightforwardly
obtained for the HMF model,
and holds in the long-range regime $0\leq\alpha<1$
due to the universality of the canonical partition function \cite{campa-00}.

Although we do not show the derivation, 
we can show the following relation for $T\geq T_{c}$, $T_c$ being the 
canonical critical temperature\cite{GY-05}
\begin{equation}
  \label{eq:deltaM-deltaT-supercritical}
  \delta_{T}(N) = [\delta_{M}(N)]^{2} + b N^{-(1-\alpha)}.
\end{equation}
The constant $b$ represents the diagonal elements of 
the matrix $(1/r_{ij}^{\alpha})$,
and can be chosen arbitrarily.
We select $b=2.0$ to obtain a fully positive spectrum
which enables us to apply the Hubburd-Stratonovitch transformation
\cite{campa-00}. 

\section{Numerical Results}

\begin{figure}[htbp]
  \centering
\includegraphics[width=8.2cm]{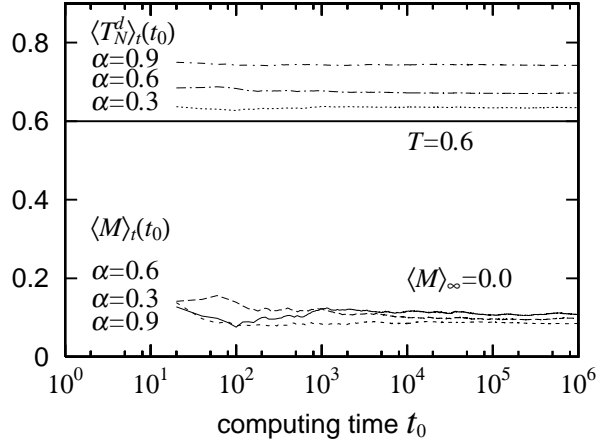}
  \caption{Time-averaged values $\langle T_N^d\rangle_{t}(t_{0})$
    and $\average{M}_{t}(t_{0})$ as functions of computing time $t_{0}$.
    Parameters are taken at $N=256$, $\alpha=0.3,0.6,0.9$ and $U=0.8$, 
    for which $\average{M}_\infty=0.0$.    
    The horizontal solid line represents the corresponding canonical 
    temperature $T=0.6$.}
  \label{fig:time-average}
\end{figure}

Let us check the validity of our theoretical predictions of $\delta_T$ and 
$\delta_M$ numerically.
The time-averaged temperature $\average{T_N^d}_{t}$
and the order parameter $\average{M}_{t}$ are defined as follows:
\begin{equation}
  \label{eq:def-T-M}
  \average{T_N^d}_{t}(t_{0}) := \dfrac{1}{t_{0}} \int_{0}^{t_{0}} 
   T_N^d(t)~dt,\quad
  \average{M}_{t}(t_{0}) := \dfrac{1}{t_{0}} \int_{0}^{t_{0}} M(t) ~dt.
\end{equation}
Numerical integrations of the canonical equation of motion
are performed by using a fourth-order symplectic integrator.
\cite{yoshida-93}.
Thanks to the Fast Fourier Transformation technique,
the calculation times are of order $N\ln N$.
The time slice of the integrator is set at $0.2$
and it suppresses the relative energy error $\Delta E/E\sim 10^{-5}$.
Our initial condition is, for all values of $\alpha$,
the canonical equilibrium distribution of the HMF model,
that is $p$ and $q$ are randomly taken from the distributions
being proportional to $\exp(-\beta p^{2}/2)$ and
$\exp(\beta \average{M}_{\infty}\cos q)$
\cite{yamaguchi-04} respectively, where $\beta=1/T$.
The sum of momenta is an integral of motion,
and we initially set it to zero.
The computing time, $t_{0}$
in Eq.\fr{eq:def-T-M},
are set to more than $10^{6}$ to realize equilibrium states
in the sense of dynamics, where we define such equilibrium
as when the time-averaged macroscopic variables are almost stational.
Typical time sequence of both 
$\average{M}_t(t_0)$ and 
$\average{T_N^d}_t(t_0)$ are 
shown in Fig.\ref{fig:time-average}
as functions of computing time $t_{0}$.
In Fig.\ref{fig:time-average}, we observe finite size effects 
$\delta_{M}$ and $\delta_{T}$,
which are discrepancies from $\average{M}_{\infty}$ and $T$ respectively.

\begin{figure}[htbp]
  \centering
\includegraphics[width=6.8cm]{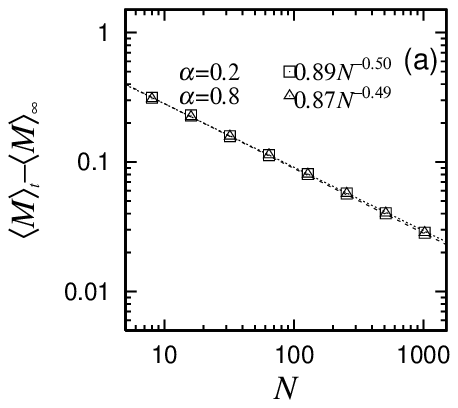}
\includegraphics[width=6.8cm]{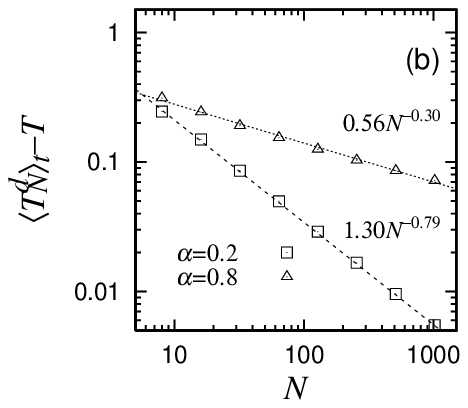}
  \caption{Dependence on $N$ of the two discrepancies
    (a) $\average{M}_{t}$-$\average{M}_{\infty}$ and
    (b) $\langle T_N^d\rangle_{t}$-$T$ at $U=5.0$.
    Asymptotic power-law decays are observed.}
  \label{fig:MT-sup}
\end{figure}

For the supercritical energy regime,
we investigate the finite size effects for two values of energy.
One is the high enough energy $U=5.0$, 
and the other one is supercritical but the middle energy $U=0.8$.
The discrepancies $\delta_M (N)$ and $\delta_{T}(N)$
are shown in Fig.\ref{fig:MT-sup}
as functions of $N$.
It is shown that the discrepancies
algebraically decrease as $N$ increases, that is
\begin{equation}
  \label{eq:power-law}
  \delta_{M}(N) \propto N^{-\mu_M(\alpha,U)},
  \quad
  \delta_{T}(N) \propto N^{-\mu_T(\alpha,U)}.
\end{equation}
According to the theoretical prediction 
\fr{eq:deltaM-deltaT-supercritical},
the exponent $\mu_{M}$ should be related to the other $\mu_{T}$ as
\begin{equation}
  \label{eq:deltaM-deltaT-exponent}
  \mu_T = \min (2\mu_M , 1-\alpha).
\end{equation}
Let us check the relation \fr{eq:deltaM-deltaT-exponent}
by observing the exponents $\mu_{M}$ and $\mu_{T}$
as functions of $\alpha$.

\begin{figure}[htbp]
  \centering
\includegraphics[width=6.8cm]{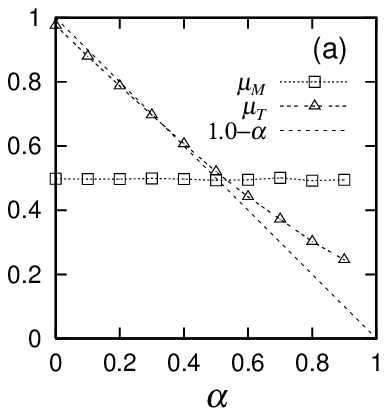}
\includegraphics[width=6.8cm]{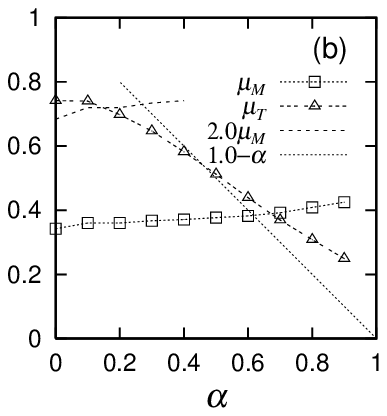}
  \caption{Exponents $\mu_M$ and $\mu_T$ as functions of $\alpha$.
    (a) $U=5.0$. (b) $U=0.8$. Dashed curves represent theoretical predictions.}
  \label{fig:exponents}
\end{figure}

For the high enough energy $U=5.0$ (see Fig.\ref{fig:exponents}(a)),
the theoretical curve is in good agreement with numerical results
in the interval $0\leq\alpha\lesssim 0.6$. 
In addition, the theoretical prediction is 
qualitatively good for $U=0.8$
(see Fig.\ref{fig:exponents}(b)).
For further discussions, we refer the reader to Ref.\citen{GY-05}. 


\section{Conclusions}

In this article 
the $\alpha$-XY model \fr{eq:alpha-XY}
with $d=1$ has been studied for the long-range regime $0\leq\alpha<1$,
so that we clarify the role of interaction length in Hamiltonian 
systems with many degrees of freedom\cite{GY-05}.
We have studied dependences on degrees of freedom
for time-averaged macroscopic variables calculated
through Hamiltonian dynamics with finite degrees of freedom,
contrasted with the canonical averages of ones 
in the thermodynamic limit. 
A non-universal behavior in $0\leq\alpha<1$ 
is observed in the asymptotic behavior
of the discrepancies for the supercritical regime.
For the supercritical regime,
the discrepancy between the two averages decays as a power-type function,
and the exponents depending on $\alpha$ 
are explained by applying the canonical ensemble
with finite degrees of freedom with the aid of the saddle-point approximation.
This non-universal behavior sheds light on the study of Hamiltonian systems
with long-range interactions. 

\section*{Acknowledgments}
One of the authors (SG) has been supported by 
a JSPS Fellowship for Young Scientists. 
SG also thanks the organizers 
for giving SG a good opportunity to talk about this study
in the conference. 
The other author (YYY) has been partially supported
by the Ministry of Education, Science, Sports and Culture,
Grant-in-Aid for Young Scientists (B), 16740223, 2004.

\section*{Appendix}
\begin{figure}[htbp]
  \centering
 \includegraphics[width=6.8cm]{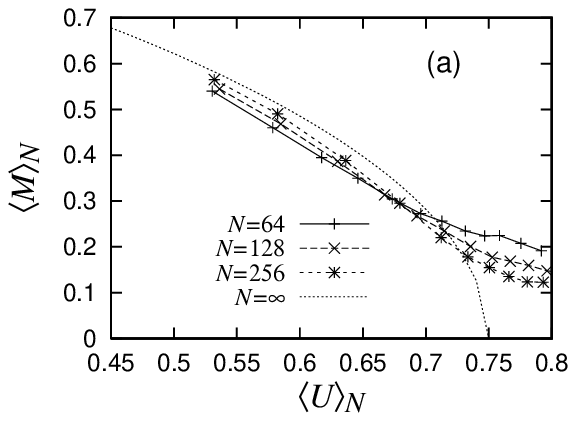}
\includegraphics[width=6.8cm]{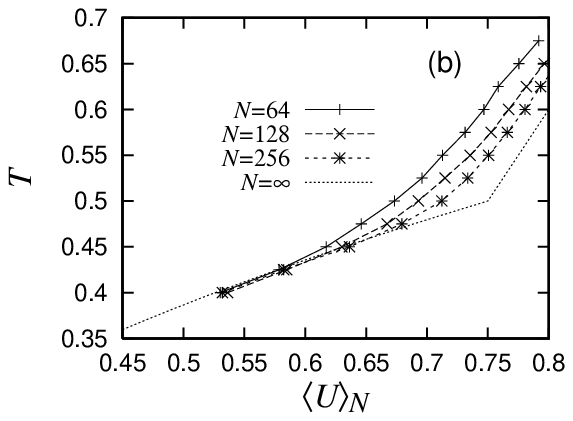}
  \caption{Dependence on $N$ of the values of macroscopic variables  
     in the canonical ensemble with $\alpha=0.3$, 
    (a) $\average{M}_{N}$-$\average{U}_{N}$ and
    (b) $T$-$\average{U}_N$.}
\label{fig:MT-mtr-a3}
\end{figure}

\begin{figure}[htbp]
  \centering
\includegraphics[width=6.8cm]{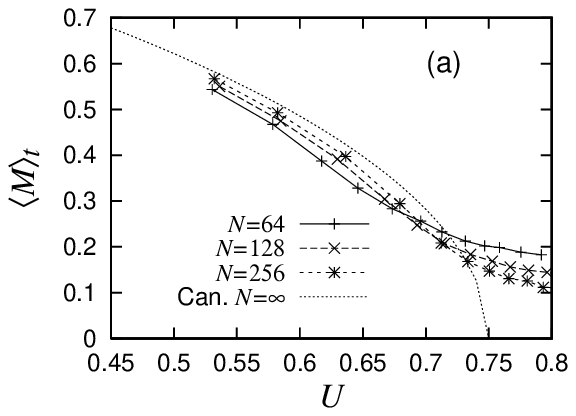}
\includegraphics[width=6.8cm]{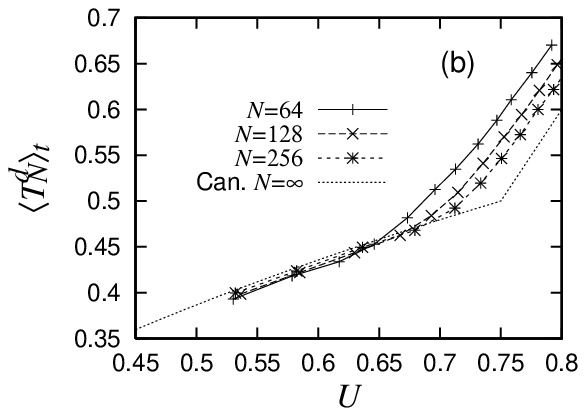}
  \caption{Dependence on $N$ of the values of macroscopic variables
     in the dynamical system with $\alpha=0.3$,
    (a) $\average{M}_{t}$-$U$ and
    (b) $\langle T_N^d\rangle_{t}$-$U$.
  Here the lines denote the canonical averages of $M$ and $T_N^d$ 
  in $N\to\infty$.}
  \label{fig:MT-dyn-a3}
\end{figure}

\begin{figure}[htbp]
  \centering
\includegraphics[width=6.8cm]{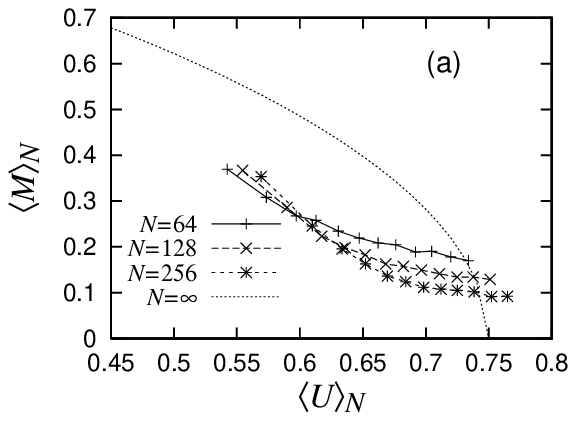}
\includegraphics[width=6.8cm]{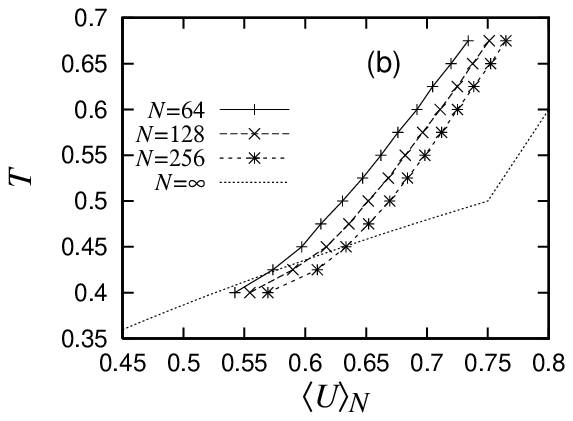}
  \caption{Dependence on $N$ of the values of macroscopic variables  
     in the canonical ensemble with $\alpha=0.9$, 
    (a) $\average{M}_{N}$-$\average{U}_{N}$ and
    (b) $T$-$\average{U}_N$.}
\label{fig:MT-mtr-a9}
\end{figure}

\begin{figure}[htbp]
  \centering
\includegraphics[width=6.8cm]{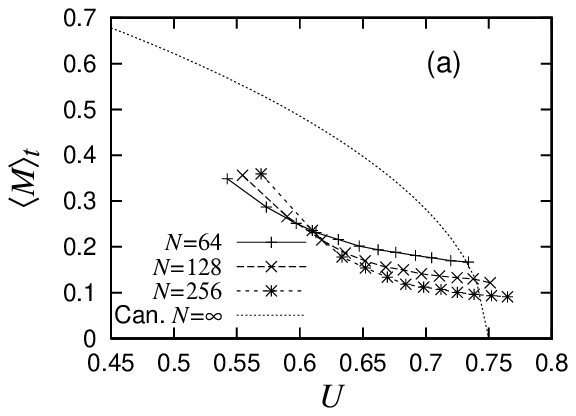}
\includegraphics[width=6.8cm]{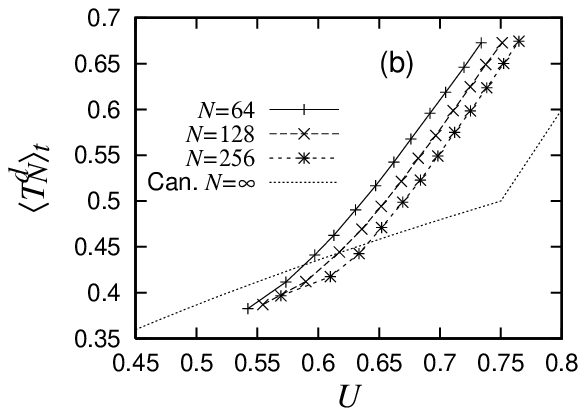}
  \caption{Dependence on $N$ of the values of macroscopic variables
     in the dynamical system with $\alpha=0.9$,
    (a) $\average{M}_{t}$-$U$ and
    (b) $\langle T_N^d\rangle_{t}$-$U$.
  Here the lines denote the canonical averages of $M$ and $T_N^d$ 
  in $N\to\infty$.}
  \label{fig:MT-dyn-a9}
\end{figure}

In this appendix, to show that 
the values of the canonical averages of macroscopic variables are 
close to the values of time-averages of ones, 
we numerically study 
equilibrium values of macroscopic variables 
for models with finite $N$ and $d=1$ in each situation.
In the canonical ensemble and in the dynamical system, 
we use a conventional canonical Monte Carlo algorithm 
and a fourth symplectic integrator respectively. 
First, for the model with $\alpha=0.3$, 
and this $\alpha$ is close to $\alpha=0$,   
Figs.\ref{fig:MT-mtr-a3} and \ref{fig:MT-dyn-a3} show that the 
relation $\average{M}_N$-$\average{U}_N$ in the canonical ensemble are 
close to $\average{M}_t$-$U$ in the dynamical system.
Furthermore, it is suggested that the time-averaged values of 
macroscopic variables go towards the canonical averages of 
ones evaluated in the thermodynamic limit as $N$ increases.
Next, 
for the model with $\alpha=0.9$, we show the values of
the magnetization and the temperature as functions of the energy density 
in Figs.\ref{fig:MT-mtr-a9} and \ref{fig:MT-dyn-a9}.
The relation $\average{M}_N$-$\average{U}_N$ in the 
canonical ensemble are close to $\average{M}_t$-$U$ 
in the dynamical system again. 
However, in this case, 
the prediction from the canonical average evaluated in the thermodynamic limit 
is not close to the relations 
$\average{M}_N$-$\average{U}_N$ and $\average{M}_t$-$U$. 

Thus, from the observations in the two cases 
$\alpha=0.3$ and $\alpha=0.9$,  
we conclude that, at a fixed value of $\alpha$ provided $\alpha<1$,
the relations $\average{M}_N$-$\average{U}_N$ in the canonical ensemble 
are close to the relations $\average{M}_t$-$U$ in the dynamical system.
Meanwhile, at any fixed value of $N$, 
the discrepancies between $\average{M}_N$-$\average{U}_N$ and 
$\average{M}_{\infty}$-$\average{U}_{\infty}$, and between $\average{M}_t$-$U$
and $\average{M}_{\infty}$-$\average{U}_{\infty}$,  become 
larger as $\alpha$ increases up to unity.
This tendency of the discrepancy is consistent with ones 
in Fig.\ref{fig:exponents} for the supercritical energy regime.



\end{document}